# Application of Integral Value Transformation (IVT) in a Specialized Computer Network Design

*Souvik Naskar, Avishek Ghosh, Pabitra Pal Choudhury*

*Applied Statistics Unit, Indian Statistical Institute, Calcutta, India*

*Email: sviknskr@gmail.com, avishek.ghosh38@gmail.com, pabitrapalchoudhury@gmail.com*

***Abstract***

*Integral Value Transformation (IVT) is a family of transformations from $N_0^k$ to $N_0$ . An algebraic result has been established in p-adic IVT systems and an application of the result is described in this paper. The result in this paper provides the rule to find the $p^{th}$ pre image of a natural number for the Collatz like bijective functions in p-adic IVT systems. Using this result a routing algorithm is proposed. This proposed routing algorithm reduces number of searches and address calculation.*

## *1. Introduction:*

Collatz conjecture or the 3x+1 problem is a well known problem [1]. The Collatz function is defined on N as,

$$f(x) = 3x + 1 \; when \; x \; is \; odd$$

$$= x/2 \; when \; x \; is \; even$$

An iterative scheme is introduced as $X_{n+1} = f(X_n)$. There is a natural number $i$ such that $X_i = 1$ for all initial value $X_0$. The sequence $\{X_n\}$ is known as Collatz sequence.

In the similar fashion one dimensional IVT from $\mathbb{N} \cup \{0\}$ $to$ $\mathbb{N} \cup \{0\}$ is defined as

$IVT^{p,1}{}_j(x) = \left(f_j(x_n) f_j(x_{n-1}) \dots \dots \dots f_j(x_1)\right)_p = m$; where m is the decimal conversion from the p adic number, and $x = (x_n \; x_{n-1} \dots \dots x_1)_p$.

In [2] Hassan et al. have proved that an iterative scheme $Y_{n+1} = IVT_1^{2,1}(Y_n)$ converges to 0 for any initial value $Y_0$. Here $\{Y_n\}$ is denoted as IVT sequence.

In this paper we mainly focus on the bijective Collatz like IVTs. For a given number, the rule for calculating its $p^{th}$ pre-image for bijective Collatz like IVTs is described. Using this rule, an efficient single path routing protocol is designed.

In practice there are many algorithms (both static and dynamic) proposed and many of them have been implemented to facilitate optimum routing namely *Hierarchical Routing* , *Fuzzy Routing, Multiple Agents algorithm, Heuristic routing, Deflection routing, Geographic routing, Multipath routing* etc[3,4,5,6 & 7]. One of most popular routing method viz. *Distance vector routing* use the *Bellman-Ford algorithm* [8]. This approach assigns a number, the cost, to each of the links between each node in the network. Nodes will send information from point A to point B via the path that results in the lowest total cost (i.e. the sum of the costs of the links between the nodes used). When applying link-state algorithms, each node uses as its fundamental data a map of the network in the form of a graph. The shortest route algorithm proposed by Dijkstra (1959) is quite famous and efficient [9].

Our proposed single path static routing protocol uses the property of Integral Value Transformations (IVT) .The algorithm uses the underlined property of Collatz like bijective IVT to optimize routing. Before coming to the technical details let us revisit the notion of Integral Value Transformation.

## 2. *Notion of IVT:*

Integral Value Transformations (IVT) $IVT_{\#}^{p,1}$ from ${\mathbb{N}_0}^{k}$ to $\mathbb{N}_0$ is defined where p denotes the p-adic number, k denotes dimension of the domain and # represents the transformation index [2]. It is worth noting that these IVTs' correspond to each of the multistate Cellular Automata.

Let us define the IVT in $\mathbb{N}_0$ in 4-adic number systems. There are 256 ($4^{4^1}$) one variable four state CA rules. Corresponding to each of those CA rules there are 256 IVTs are there in 4 adic systems in one dimension.

${\mathrm{IVT}^{4,1}}_{\#}$is mapping a non-negative integer to a non-negative integer.

$${\mathrm{IVT}^{4,1}}_{\#}(\mathrm{a}) = \left((\mathrm{f}_{\#}(\mathrm{a_n})\mathrm{f}_{\#}(\mathrm{a_{n-1}}) \ldots \mathrm{f}_{\#}(\mathrm{a_1})\right)_4 = b$$

Where 'a' is a non-negative integer and $a = (a_n a_{n-1} \ldots a_1)_4$ and 'b' is the decimal value corresponding to the 4-adic number.

For an example, let us consider $a = 433 = (12301)_4$ and $\# = 114$ so $\mathrm{f}_{\#}(0) = 2;\ \mathrm{f}_{\#}(1) = 0;\ \mathrm{f}_{\#}(2) = 3\ and\ \mathrm{f}_{\#}(3) = 1$.

Therefore, ${\mathrm{IVT}^{4,1}}_{114}(772) = (\mathrm{f}_{114}(1)\mathrm{f}_{114}(2)\mathrm{f}_{114}(3)\mathrm{f}_{114}(0)\mathrm{f}_{114}(1))_4 = (03120)_4 = 216$.

Consequently, ${\mathrm{IVT}^{4,1}}_{114}(433) = 216$.

In the next section we will discuss about a class of IVT, which are bijective and Collatz like in nature.

### *2.1. Bijective Collatz like IVTs*

**Theorem2.1** *There are $(p-1)!$ number of bijective functions in the set of Collatz functions in p-adic system.*

**Proof:**

It is clear that all the $p^p$ functions are surjective but all of them are not injective. A bijective function has all different entries (0, 1, 2, …$p-1$) corresponding to different arguments (0, 1, 2,…$p-1$) of the function. It will be Collatz like if there exists a path to reach 0 (attractor in this case) for each argument (0, 1, 2, …$p-1$). It is clear that for Collatz like bijective functions, 0 can't map to itself (because the path to attractor seizes if 0 maps to itself). So 0 can map to other $p-1$ number of arguments. For each argument, other $p-2$ arguments can combine themselves in $(p-2)!$ ways. In each of this case there exists a path to reach the attractor 0. So the total number of cases where a path exists to the attractor is $(p-1)(p-2)!$ i.e. $(p-1)!$. Hence the total number of collatz like bijective functions is $(p-1)!$.

***Corollary:*** There are $p^p$ number of functions in $p$ adic IVT system and number of bijective functions is $p!$ [10]. It can be easily shown that out of them $p^{p-1}$ number of functions are Collatz-like [11]. Among the non-collatz functions, $(p-1)(p-1)!$ number of functions are bijective.

Let us denote the set of all Collatz like bijective IVTs as $\Gamma^{p,1}$. For a particular function belonging to $\Gamma^{p,1}$ , let the mapping $f_{\#}(i) = j$ (where $0 \leq i, j \leq p-1$) holds. For a bijective Collatz like function, $i \neq j$. So we can say that the functional rule forms a cycle, indicating that by applying the rule on a $d_i$ (where $0 \leq d_i \leq p-1$), we can achieve $d_i$ again after p iterations.

For example,

In 4-adic number system consider # =114. $114_{10}=1302_4$. For # =114, $f_{114}(0) = 2$, $f_{114}(1) = 0$, $f_{114}(2) = 3$ and $f_{114}(3) = 1$.

These function forms a cycle shown below:

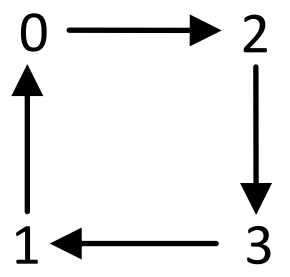

Taking a natural number A=56913. Applying $IVT_{114}^{4,1}$ on 56913 iteratively 4 times, the results are:

$$56913_{10} = 31321101_4$$
$$18184_{10} = 10130020_4$$
$$8622_{10} = 2012232_4$$
$$14583_{10} = 3203313_4$$
$$7761_{10} = 1321101_4$$

Starting from the LSB to the left, it is observed that the digits are repeated at the 4th iteration up to a certain position. As 3 at the MSB becomes 0 in the second iteration, it is omitted.

This property is utilized to find the $p^{th}$ pre-image of a natural number. This is described in the next section.

## *3. $P^{th}$ pre-image of a natural number:*

***Theorem 3.1.*** *For every natural number N, there exists A (A>N) such that* $\left(IVT_{\#}^{p,1}\right)^p(A) = N$.

*Where,*
$$A = N + \left(p^{|N|+1}\right) \times \gamma, \text{ if } IVT_{\#}^{p,1}(0) = \alpha, IVT_{\#}^{p,1}(\gamma) = 0$$
$$= N + \left(p^{|N|}\right) \times \gamma, \text{ if } IVT_{\#}^{p,1}(\gamma) = \alpha$$

*α is MSD of N in p-adic representation.* $|N|$ *is number of digits in in p-adic representation of N* , $0 < \gamma \leq p-1$ *and* $IVT_{\#}^{p,1} \in \Gamma^{p,1}$.

**Proof:** As $\Gamma^{p,1}$ consists of bijective and Collatz like IVTs, it is obvious that there will be a cycle formed by $f_{\#}$ corresponding to $IVT_{\#}^{p,1}$ . Consequently, $\left(IVT_{\#}^{p,1}\right)^p(n) = n$ where $0 \leq n \leq p-1$.

It is obvious that A must be a natural number. Let the p-adic representation of N is $(\alpha\ N_{n-2} \ldots\ldots N_0)_p$.

Consider $A = (\gamma\ 0\ \alpha\ N_{n-2} \ldots\ldots N_0)_p = \gamma \times p^{n+1} + 0 \times p^n + \alpha \times p^{n-1} + N_{n-2} \times p^{n-2} + \cdots\ldots N_0 \times p^0 = N + (p^{|N|+1}) \times \gamma$.

Next it remains to be shown that $\left( IVT_{\#}^{p,1}(A) \right)^p = N$.

Since $IVT_{\#}^{p,1}(\gamma) = 0$ and $\left( IVT_{\#}^{p,1}(n) \right)^p = n\ \forall\ 0 < n \leq p-1$

$$\left( IVT_{\#}^{p,1}(A) \right)^p = N.$$

For illustration of 3.1, $(\mathrm{IVT}_{57}^{4,1})^4(12774) = 486$, i.e. N=486, A=12774.

In the second case, where $IVT_{\#}^{p,1}(\gamma) = \alpha$, let the p-adic representation of A be,

$$A = (\gamma\ \alpha\ N_{n-2} \ldots\ldots N_0)_p = \gamma \times p^n + \alpha \times p^{n-1} + N_{n-2} \times p^{n-2} + \cdots\ldots N_0 \times p^0 = N + (p^{|N|}) \times \gamma$$

As $\gamma$ produces 0 after $\alpha$, and $\left( IVT_{\#}^{p,1}(n) \right)^p = n\ \forall\ 0 < n \leq p-1$

$$\left( IVT_{\#}^{p,1}(A) \right)^p = N.$$

For illustration of 3.2, $(\mathrm{IVT}_{57}^{4,1})^4(1766) = 742$, i.e. N=742, A=1766.

## 4. *Application:*

The fundamental function of the network layer is to provide routes for packets (in case of packet switching) from the source machine to the destination machine. In most cases, packets will require multiple hops to make the journey. In our protocol we used the property of bijective Collatz like IVTs and the rules we established in section 3.

### *4.1. Proposed Protocol:*

Consider the situation where a number of source machines are trying to send data to a single destination machine or target machine.

Each source machine is given a natural number as an address. The target machine is designated by the number 0. The routing protocol is described below.

Each source machine sends data to the immediate machine with address $IVT_{\#}^{p,1}(N)$ where $N$ the address of the source machine is and # denotes bijective collatz like IVT function in p adic domain.

The source machine will store the number $IVT_{\#}^{p,1}(N)$ along with some information regarding the machine with address$IVT_{\#}^{p,1}(N)$, so that for further communications it will not compute or search for the machine to send its data. It will straightly forward data to $IVT_{\#}^{p,1}(N)$. So a virtual link (path) is established between machines $N$ and $IVT_{\#}^{p,1}(N)$.

As the function # is collatz like, it is assured that there exists a path from any source machine to the target machine as the destination machine is designated by 0 which is the attractor of the Collatz like IVT functions.

With the help of above described protocol, assume machine M sends data to destination machine in finite number of hops. So a virtual link is established between M and destination machine. For a given M, a natural number A (A>M) can be found whose trajectory includes M. A can be calculated using the result of *Theorem 3.1*. For simplicity, we consider $A = \varphi(M)$.When A tries to communicate and send its data to the destination, after p iterations (hops) it reaches M. Once it reaches M, the path from M to destination is already established in the sense that no computing and searching is needed to route data from M to the target machine. So a part of the path from A to destination is pre-calculated (if M sends data prior to A) which leads to the optimization in routing.

Now another case is considered where A has communicated prior to M. In this case the path from A to destination is virtually established and M belongs to that path. So a path exists from M to destination and the communication from M to the target machine no longer needs any computation or searching.

### *4.2. Design:*

The designer must be aware that how many paths will be there which are directly connected to the destination node or in other words how many parallel paths are there in the network. We take the number as $v$. We name these nodes as $M_1$, $M_2$, $M_3$....$M_v$. The number of these nodes is chosen in such a way that $IVT_{\#}^{p,1}(M)$ produce 0 in the first iteration. So it is easy to see that from these nodes the destination nodes can be accessed in one step. Any other nodes will be successively connected to these first nodes. For designing the topology of the network, we first calculate the $p$ th pre-image of a node directly connected to the destination node. If the number of nodes required for a level is greater than p, then we take further pre-images so that the number of nodes can be accommodated within the design. As an example, if the designer chose p=5 and a level needs 10 nodes, then we take the $10^{th}$ (5 X 2) pre image of the first node using the φ(M) function twice. Now all of the nodes can be accommodated.

So, for designing the network, three parameters are used: the number of machines that are connected to the destination node, denoted by $v$, the number of phases in each parallel line, $n$, and the $p$ for which the $IVT_{\#}^{p,1}$ will be computed. Using these parameters, we can accommodate at most $(n \times p + 1) \times v + 1$ number of nodes. There are three parameters in this equation and they can be chosen according to the requirement of the network, available technology, cost etc. The choice of p depends on the total number of nodes in the network. The designer may chose the value of n and v in such a way that the value of p doesn't exceed 5. By doing so, we not only make the network design feasible but also reduce the number of unused nodes. In most cases number of parallel paths may be restricted by the network technology, so $v$ is restricted. The parameter $n$, which denotes the number of phases in the network is unrestricted and for large n, the number of optimal pairs is increased.

As there are (p-1)! numbers of Collatz like bijective functions for a p-adic system; it is easy to see that there will be (p-1)! different allocations of nodes. But if for each level, the number of nodes is the same, the network will be *isomorphic*. So by changing the value of #, we can change the allocation of the nodes but not the topology of the graph. As an example, for p=4 and # =57, the network looks like this:

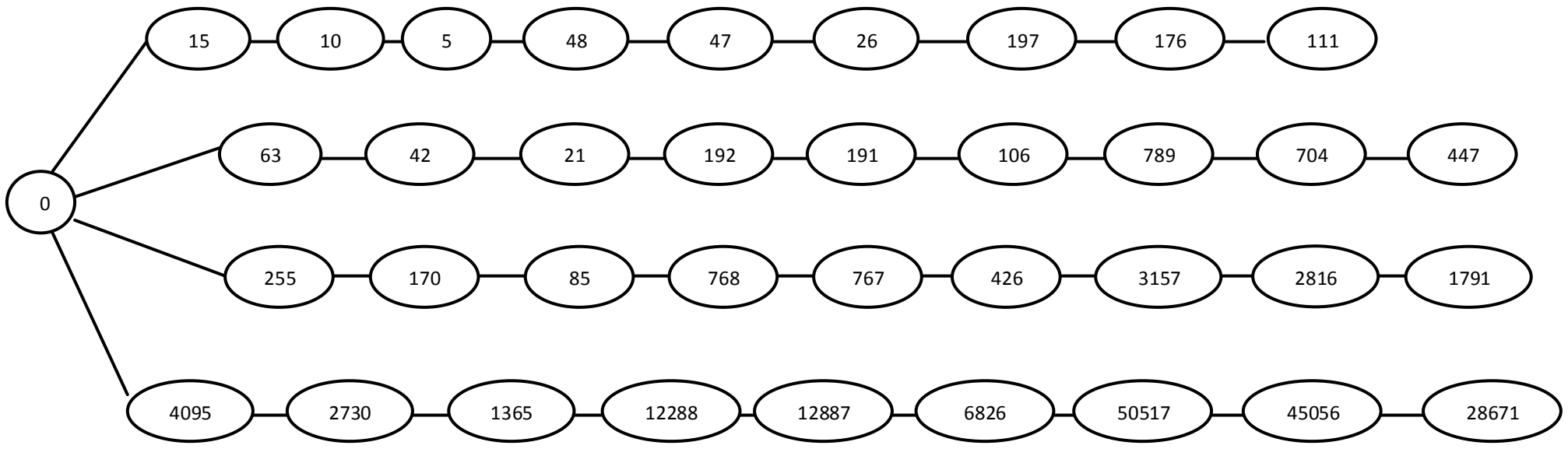

*Figure 1. Network topology for* p=4 and # =57,

In Figure 1, (15, 47), (10, 26), (5, 197), (48, 176), (47, 111) are optimal node-pairs in the first parallel path. Similarly optimal pairs can be determined for each parallel path.

And for p=4 and # =114 the same network will look like this

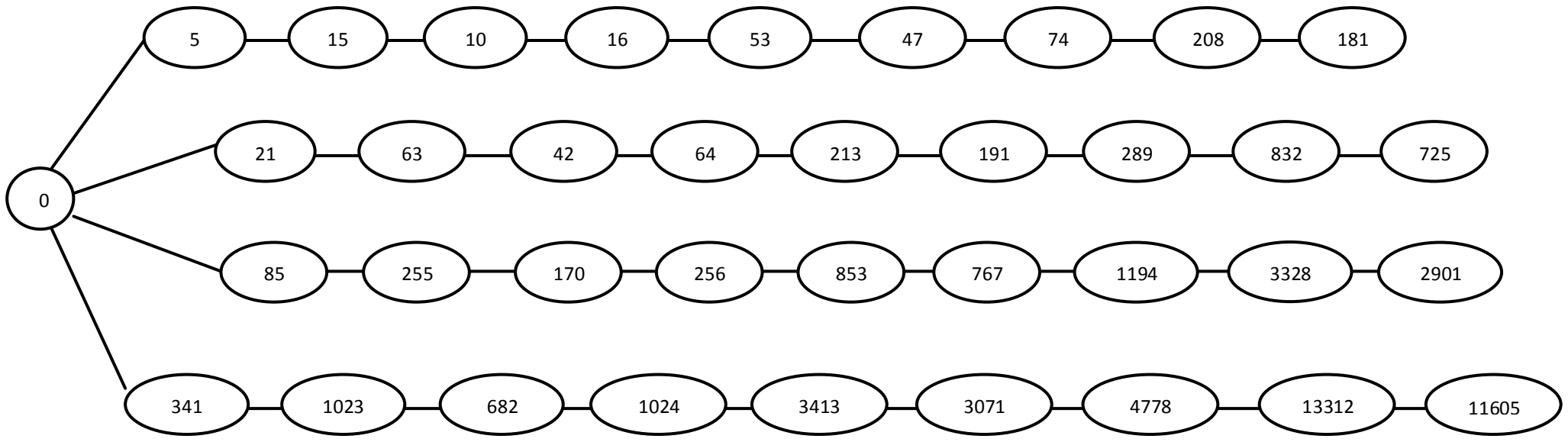

*Figure 2. Network topology for* p=4 and #=114,

Here we can observe an important property of the network. Since we are using Collatz like bijective IVT systems, we can be assured that there is only one path between two particular nodes. This implies that if there is a path between two nodes, these two nodes form a pair. And there exist a number of pairs in the network.

## 5. Conclusion and future works:

In this paper an algebraic result of finding the $p^{th}$ pre-image of a number for Collatz like bijective IVTs is explored. Using it an efficient routing protocol is designed. A further extension of the protocol can be done by varying the number of nodes in a parallel line or using variable p for each parallel line. Dynamic address allotment to the source machines and the scheduling methodologies would be our future endeavors.

## Acknowledgement

Authors would like to acknowledge their colleague, Sk. Sarif Hassan for his technical help.